\documentclass[aoas,preprint]{imsart}

\RequirePackage[OT1]{fontenc}
\RequirePackage{amsthm,amsmath}
\RequirePackage{natbib}
\RequirePackage[colorlinks,citecolor=blue,urlcolor=blue]{hyperref}
\usepackage{graphics,url}
\usepackage{pstricks,pst-node,pst-tree}
\usepackage{amsmath,longtable}
\usepackage{lscape}
\startlocaldefs
\numberwithin{equation}{section}
\theoremstyle{plain}

\endlocaldefs

\begin{document}

\begin{frontmatter}
  \title{Classification and regression tree methods for
    incomplete data from sample surveys
  }

  \runtitle{Trees for incomplete data}
  

\begin{aug}
  \author{\fnms{Wei-Yin} \snm{Loh}\thanksref{m1,t1}
    \ead[label=e1]{loh@stat.wisc.edu}},

  \author{\fnms{John} \snm{Eltinge}\thanksref{m2}
    \ead[label=e2]{Eltinge.John@bls.gov}},

  \author{\fnms{MoonJung} \snm{Cho}\thanksref{m2}
    \ead[label=e3]{Cho.Moon@bls.gov}}

  \and
  
  \author{\fnms{Yuanzhi} \snm{Li}\thanksref{m1}
    \ead[label=e4]{yuanzhi.li@wisc.edu}
    \ead[label=u1,url]{http://www.foo.com}}

  \thankstext{t1}{Supported in part by NSF grant
    DMS-1305725 and an ASA/NSF/BLS research fellowship.}

  \runauthor{W.-Y. Loh et al.}

  \affiliation{University of Wisconsin-Madison\thanksmark{m1} and Bureau of
    Labor Statistics\thanksmark{m2}}

\address{Department of Statistics\\
  University of Wisconsin\\
  1300 University Avenue\\
  Madison, WI 53706\\
  \printead{e1}\\
  \printead{e4}}

\address{Office of Survey Methods Research\\
  Bureau of Labor Statistics\\
  2 Massachusetts Avenue, NE\\
  Washington, DC 20212\\
  \printead{e2}\\
  \printead{e3}}
\end{aug}

\begin{abstract}
  Analysis of sample survey data often requires adjustments to account
  for missing data in the outcome variables of principal interest.
  Standard adjustment methods based on item imputation or on
  propensity weighting factors rely heavily on the availability of
  auxiliary variables for both responding and non-responding units.
  Application of these adjustment methods can be especially
  challenging in cases for which the auxiliary variables are numerous
  and are themselves subject to substantial incomplete-data problems.
  This paper shows how classification and regression trees and forests
  can overcome some of the computational difficulties.  An in-depth
  simulation study based on incomplete-data patterns encountered in
  the U.S. Consumer Expenditure Survey is used to compare the methods
  with two standard methods for estimating a population mean in terms
  of bias, mean squared error, computational speed and number of
  variables that can be analyzed.
\end{abstract}



\end{frontmatter}

\section{Introduction}
\label{sec:intro}

\nocite{raghu04} We consider estimation of a population mean $\mu$ of
a variable $Y$ in simple random sampling without replacement from a
finite population when the sample $S$ is incompletely observed.  We
first review several existing solutions and then propose some new
solutions based on classification and regression trees and forests. We
aim to show, by means of a realistic simulation study, that certain
tree methods are as good as or better than two standard methods in
four important respects: (i)~bias, (ii)~mean squared error,
(iii)~computational speed, and (iv)~applicability to large numbers of
prospective predictor variables.  Within the large and complex space
of incomplete-data methods for surveys, one does not expect any given
computational procedure to dominate all other procedures uniformly
with respect to each of criteria (i)--(iv).  Instead, performance will
depend on several factors, including (a)~the number of prospective
predictor variables, and the degree of association among those
variables; (b)~incomplete-data patterns encountered among the
predictor variables; and (c)~the extent to which the observed data and
missing-data patterns are consistent with model conditions that are
used implicitly or explicitly by a given computational procedure.

Let $n$ denote the number of subjects in $S$ and let $S_1 \subset S$
be the subset containing the non-missing $Y$ values. Let $y_i$ denote
the value of $Y$ for subject~$i$ in $S_1$.  If the probability $\pi_i$
that $Y$ is non-missing for subject $i$ is known, then
$n^{-1} \sum_{i \in S_1} \pi_i^{-1} y_i$ is an unbiased estimate of
$\mu$ (throughout this paper, expectations are evaluated with respect
to both the sample design and the nonresponse mechanism).  Let
$\hat{\pi}_i$ be an estimate of $\pi_i$ if the latter is unknown. The
\emph{inverse probability weighted} (IPW) estimate of $\mu$ is
\citep[see, e.g.,][]{little86,SW13}
\begin{equation}
  \Big(\sum_{i \in S_1} \hat{\pi}_i^{-1} \Big)^{-1} \sum_{i \in S_1} \hat{\pi}_i^{-1} y_i.
  \label{eq:ipw}
\end{equation}
The bias of the IPW estimate depends on accurate specification and
estimation of a model for $\pi_i$.  In applications where covariates
are completely observed, logistic regression is typically used. The
sample mean, $\bar{y} = n_1^{-1} \sum_{i \in S_1} y_i$, where $n_1$ is
the number of observations in $S_1$, is an IPW estimate with
$\hat{\pi}_i = n_1/n$ for $i \in S_1$.

An alternative approach is \emph{imputation} of the missing values.
Let $S_2 = S - S_1$ be the subset of the sample with missing $Y$.  Let
$\hat{y}_j$ denote the imputed value of $Y$ for subject $j$ in $S_2$.
Then $\mu$ is estimated by the mean of the completed sample
\begin{equation}
  n^{-1} \Big(\sum_{i \in S_1} y_i + \sum_{j \in S_2} \hat{y}_j \Big).
  \label{eq:imp}
\end{equation}
The mean $\bar{y}$ of the non-missing $Y$ values is a special case of
\emph{mean imputation} with $\hat{y}_j = \bar{y}$ in (\ref{eq:imp}).
If predictor ($X$) variables are also observed, the $\hat{y}_j$ may be
obtained by \emph{regression imputation} \citep{Buck60}, where a
regression model of $Y$ on $X$ is fitted to the observations in $S_1$
and the $\hat{y}_j$ are predicted from the $X$ values in $S_2$.  If
the $X$ variables have missing values as well, the
\emph{complete-case} method fits the regression model to the subset of
values with complete observations in the $X$ and $Y$ variables.

\emph{Hot deck} \citep{LR87} methods impute missing values by random
sampling of non-missing values within `adjustment cells', which are
prespecified partitions of the data.  One way to construct the cells
is to split the range of each $X$ variable into a small number of sets
and use Cartesian products of the sets to define the cells. It can
produce, however, cells with few or no observations.  For example, in
one analysis of the Current Population Survey, the U.S. Census Bureau
had 11,232 cells constructed from seven $X$ variables \citep{boll06}.
Alternatively, the cells can be defined by partitioning the sample
according to the estimated probability that $Y$ is missing, where
these probabilities are estimated through the use of logistic
regression \citep[see, e.g.,][]{little86}. A difficulty is the choice
of $X$ variables for logistic regression if there are many $X$ or they
have missing values too.

Another method is \emph{maximum likelihood}, which draws random
observations from a parametric model fitted to the $(X, Y)$
observations.  Assuming that (i)~the parametric model is correct,
(ii)~the $X$ variables are completely observed, and (iii)~the $Y$
values are \emph{missing at random} (MAR), that is, the probability
that a value is missing does not depend on the value itself,
conditional on the non-missing values of the $X$ variables,
\citet{rubin87} showed that inferences from multiply imputed data are
statistically valid for large samples.  If there are missing $X$
values, the EM algorithm \citep{EM77} is often used to estimate the
parameters in the model. In that case, there is no guarantee that the
results are statistically valid.  AMELIA \citep{amelia} uses
multivariate normal likelihoods. Each categorical variable is first
converted to a dummy 0-1 vector and then a multivariate normal model
is fitted to all the variables. As a result, the minimum sample size
for AMELIA depends on the number of ordinal variables and the number
of dummy variables.

Yet another method is \emph{sequential regression} \citep[see,
e.g.,][]{raghubk}.  It is a regression switching technique in which
the missing values in the $X$ and $Y$ variables are initially imputed
by their means, medians, or modes. Then each variable is regressed in
turn on the other variables and missing values are updated with the
predicted values. The procedure is continued for several cycles to
reduce the effects of the initial imputed values.  MICE \citep{mice},
which stands for \emph{multiple imputation by chained equations}, is
one implementation. It uses linear regression for imputation of
ordinal variables and polytomous logistic regression for categorical
variables.  Theoretical arguments for the effectiveness of sequential
regression have been given, but they are based on the restrictive
assumption of a correct linear regression model relationship between
the variable being imputed and the covariates \citep{WC10}.

In practice, MICE can experience computational problems when there are
many predictor variables with missing values.  For example, linear
regression can fail if there is multicollinearity and logistic
regression fails if there is quasi-complete separation in the data
\citep{albert84,HLSbk}.  To overcome these problems,
\citet{burgette10} proposed replacing linear regression and logistic
regression with CART \citep{cart} classification and regression trees
from the R \texttt{tree} package. Calling the algorithm CART-MICE,
they compared it to MICE in a simulation experiment by generating
observations from a quadratic regression model with 10 correlated and
normally distributed $X$ variables, with 2 of the latter completely
observed. Missing values in $Y$ and the other 8 $X$ variables were
simulated using an MAR mechanism that depended only on the 2
completely observed $X$ variables. Their results showed that the mean
squared error and bias of the estimated regression coefficients from
CART-MICE were better than those from MICE.

\citet{rubin87} proposed imputing the missing values multiple times to
obtain variance estimates.  Although there are many simulation studies
on multiple imputation (MI), almost all used normally distributed data
and missingness mechanisms defined by linear logistic regression
models \citep[see,
e.g.,][]{allison00,schafer02,carpenter07,burgette10,WC10}.  Little is
known about the performance of the methods in real-world settings
where variables are not normally distributed (e.g., categorical
variables) and probabilities of missingness are not determined by
linear logistic regression.

To our knowledge, there are only three simulation studies that involve
real data.  \citet{YBR07} used as their simulation population a data
set with completely observed values on eight $X$ and five $Y$
variables from 1060 participants in a clinical trial. Simulated
samples of size 500 were drawn from the population by bootstrap
sampling (i.e., sampling with replacement) and missing values in the
dependent variables were artificially generated under a MAR
assumption. There were no missing values in the $X$ variables.  They
compared several MI implementations in standard statistical packages
and found that the relative performance of the methods depended on the
skewness and proportion of zeros of the $Y$ distribution; performance
was similar across methods when $Y$ deviated only slightly from
normal.

\citet{AOR07} used data from 20,378 patients in a clinical database.
The $Y$ variable was binary and there were 16 $X$ variables with
varying amounts of missing values. Missing values in the database were
imputed by MICE to produce a population without missing values. Then
samples were drawn without replacement from the population. Missing
$Y$ values in the samples were generated randomly according to a
logistic regression model fitted to the population. Finally missing
patterns in $X$ were randomly generated with probabilities
proportional to the frequencies of the top 16 missing patterns in the
original data. The authors found that MICE was one of the best
performers.

\citet{andridge10} took data from a health and nutrition survey of
16,739 respondents with completely observed values on a $Y$ variable
and seven $X$ variables as their simulation population. Simple random
samples of size 500 were drawn from the population. A logistic
regression model with three $X$ variables was used to simulate missing
values in the $Y$ variable.  They compared several variations of hot
deck imputation with MICE.

These three studies are limited by having fewer than 20 $X$ variables
and, except for \citet{AOR07}, having no missing $X$ values.  In this
article, we use a data set from the Consumer Expenditure Survey with
hundreds of variables and substantial amounts of missing values.  We
study several new methods based on the CART and GUIDE
\citep{guide,guidec,lz13} classification and regression algorithms and
compare them with AMELIA and MICE in terms of bias, mean squared error
and computational speed via simulation.  The $X$ values are
intentionally designed to be naturally missing; i.e., they are not
artificially made missing as in previous studies and hence are not
necessarily MAR. Only the $Y$ variable is MAR. No previous simulation
study comes close in terms of the number of $X$ variables and the
real-world authenticity of their missingness mechanisms.

\section{Consumer expenditure data}
\label{sec:data}
The Consumer Expenditure (CE) Quarterly Interview Survey is a
longitudinal survey sponsored by the Bureau of Labor Statistics. It
collects information on consumers' expenditures and incomes as well as
characteristics of the consumers. We use a subset of public-use
microdata of the 2013 CE Survey. Answers from 25,822 consumer units
(CUs) were obtained on more than 600 questions.  For general
background and further details on the survey, see
\citet[chap.~6]{BLSbk}.

The data contain flag variables that explain whether the values of
their associated variables are observed, missing, top-coded, etc.
Variables with underscores at the end of their names are typically
flag variables.  For example, INTRDVX\_ is the flag variable
associated with INTDRVX, the amount received by a CU in interest or
dividends during the past 12 months. Table~\ref{tab:flagcodes} gives
the possible values taken by the flag variables. In this article, we
use INTRDVX as the $Y$ variable and omit CUs with INTRDVX\_ = A and T.
This leaves a sample size of 4609 of which 1771 CUs have INTRDVX\_ =
C, i.e., INTRDVX is missing.  We ignore the sampling weight
here because the R software used in our simulations for AMELIA
\citep{amelia} and MICE \citep{mice} do not allow sampling
weights. Besides, the weights do not vary greatly in the CE data; for
example, their coefficient of variation is 0.375.

There are 630 $X$ variables that can be used to predict $Y$.
Tables~\ref{tab:19var}--\ref{tab:others} give definitions of those
used in the analyses below; see \citet{BLSdatadict} for definitions of
the other variables.  Almost 20\% (124) of the $X$ variables have
missing values. Table~\ref{tab:missinfo} lists their names and numbers
of missing values; 67 variables have more than 95\% of their values
missing (2 are missing all values).

\begin{table}
  \caption{Codes and definitions of missing value flag variables}
  \label{tab:flagcodes}
  \centering \vspace{0.5em}
  \begin{tabular}{cl} \hline
    A & valid nonresponse: a response is not anticipated \\
    C & ``don't know'', refusal or other type of nonresponse \\
    D & valid data value \\
    T & topcoding applied to value \\ \hline
  \end{tabular}
\end{table}

Note especially that these predictor variables include both variables
defined at the Consumer Unit level (e.g., housing tenure) and
variables defined for geographical areas.  The latter variable type
includes State, Region and PSU. The State identifier is subject to
confidentiality restrictions in some cases.  PSUs are small clusters
of counties. Only ``A'' size PSUs are identified and other PSU labels
are coded as missing in the CE public dataset due to confidentiality
concerns. In 2013, each ``A'' size PSU is a cluster with a population
of over 2.7 million people, and is self-representing under the CE
design.  Consequently, this paper treats ``PSU'' membership as a fixed
effect, instead of random effect, for missing value imputation.

\section{Methods}
\label{sec:methods}

We study the following ten methods on their ability to estimate the
population mean $\mu$ of INTRDVX accurately and quickly.
\begin{description}
\item[AME.]  This is AMELIA \citep{amelia} with default parameters
  except that the empirical prior level is set at 5.  According to the
  manual, the prior shrinks the covariances of the data but keeps the
  means and variances the same. It helps when there are many missing
  values, small sample sizes, large correlations among the variables,
  or categorical variables with many levels.

\item[MICE.] This is the R software \citep{mice} with default options,
  including five multiple imputations.  Problems due to
  multi-collinearity in linear regression and quasi-complete
  separation in logistic regression severely limits the number of
  variables it can employ.  With the help of previous analyses with
  similar types of data, we identified 19 variables that do not cause
  the software to fail.  Their names and numbers of missing values are
  given in Table~\ref{tab:19var}. None are income variables. Only 5 of
  the 19 have missing values (and only 1 with a significant amount).
  This set of 19 is quite easy for all imputation methods.
  
\item[SIM.] This is the simple method that estimates $\mu$ with the
  mean $\bar{y}$ of the non-missing $Y$ values, ignoring the values of
  the $X$ variables.

\begin{figure}
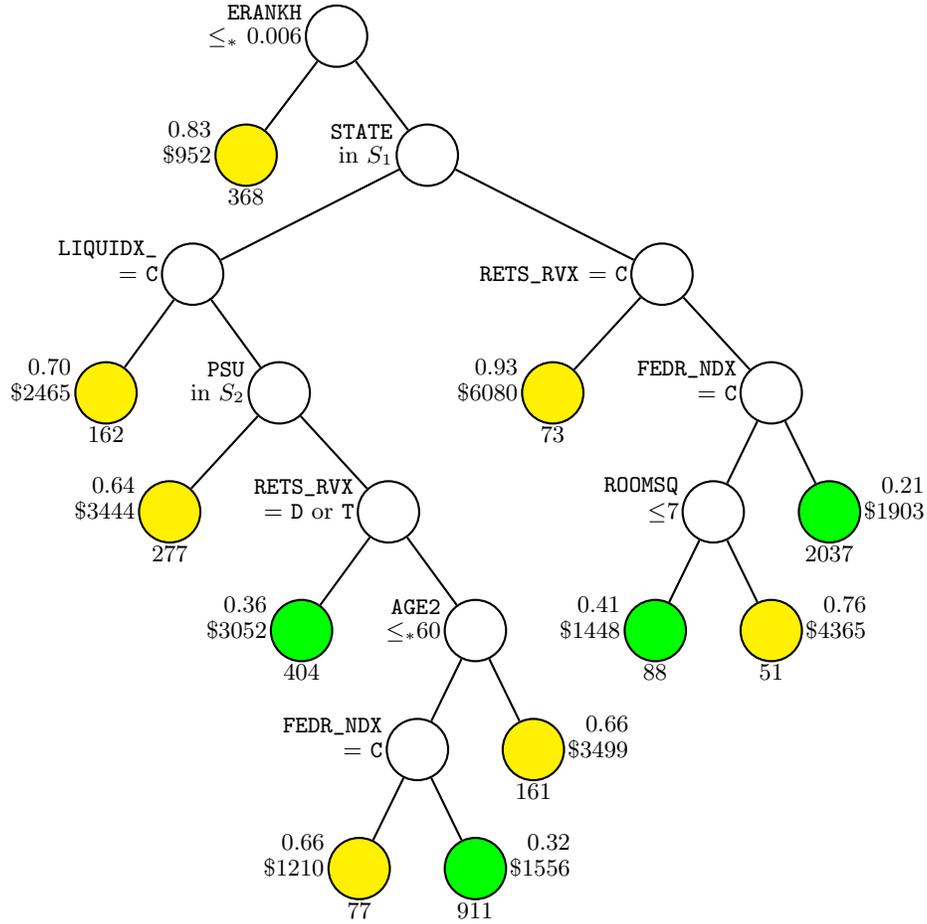

  \centering
  \psset{linecolor=black,tnsep=1pt,tndepth=0cm,tnheight=0cm,treesep=0.7cm,levelsep=45pt,radius=12pt}
  \pstree[treemode=D]{\TC~[tnpos=l]{\shortstack[r]{\texttt{\detokenize{ERANKH}}\\$\leq_*$  0.006}}}{
    \TC[fillcolor=yellow,fillstyle=solid]~{\makebox[0pt][c]{368}}
    ~[tnpos=l]{\shortstack[r]{0.83\\\$952}}
    \pstree[treemode=D]{\TC~[tnpos=l]{\shortstack[r]{\texttt{\detokenize{STATE}}\\in $S_{1}$}}}{
      \pstree[treemode=D]{\TC~[tnpos=l]{\shortstack[r]{\texttt{\detokenize{LIQUIDX_}}\\= \texttt{\detokenize{C}}}}}{
        \TC[fillcolor=yellow,fillstyle=solid]~{\makebox[0pt][c]{162}}
        ~[tnpos=l]{\shortstack[r]{0.70\\\$2465}}
        \pstree[treemode=D]{\TC~[tnpos=l]{\shortstack[r]{\texttt{\detokenize{PSU}}\\in $S_{2}$}}}{
          \TC[fillcolor=yellow,fillstyle=solid]~{\makebox[0pt][c]{277}}
          ~[tnpos=l]{\shortstack[r]{0.64\\\$3444}}
          \pstree[treemode=D]{\TC~[tnpos=l]{\shortstack[r]{\texttt{\detokenize{RETS_RVX}}\\= \texttt{D} or \texttt{T}}}}{
            \TC[fillcolor=green,fillstyle=solid]~{\makebox[0pt][c]{404}}
            ~[tnpos=l]{\shortstack[r]{0.36\\\$3052}}
            \pstree[treemode=D]{\TC~[tnpos=l]{\shortstack[r]{\texttt{\detokenize{AGE2}}\\$\leq_*$60}}}{
              \pstree[treemode=D]{\TC~[tnpos=l]{\shortstack[r]{\texttt{\detokenize{FEDR_NDX}}\\= \texttt{\detokenize{C}}}}}{
                \TC[fillcolor=yellow,fillstyle=solid]~{\makebox[0pt][c]{77}}
                ~[tnpos=l]{\shortstack[r]{0.66\\\$1210}}
                \TC[fillcolor=green,fillstyle=solid]~{\makebox[0pt][c]{911}}
                ~[tnpos=r]{\shortstack[r]{0.32\\\$1556}}
              }
              \TC[fillcolor=yellow,fillstyle=solid]~{\makebox[0pt][c]{161}}
              ~[tnpos=r]{\shortstack[r]{0.66\\\$3499}}
            }
          }
        }
      }
      \pstree[treemode=D]{\TC~[tnpos=l]{\texttt{\detokenize{RETS_RVX}} =
          \texttt{\detokenize{C}}}}{
        \TC[fillcolor=yellow,fillstyle=solid]~{\makebox[0pt][c]{73}}
        ~[tnpos=l]{\shortstack[r]{0.93\\\$6080}}
        \pstree[treemode=D]{\TC~[tnpos=l]{\shortstack[r]{\texttt{\detokenize{FEDR_NDX}}\\= \texttt{\detokenize{C}}}}}{
          \pstree[treemode=D]{\TC~[tnpos=l]{\shortstack[r]{\texttt{\detokenize{ROOMSQ}}\\$\leq$7}}}{
            \TC[fillcolor=green,fillstyle=solid]~{\makebox[0pt][c]{88}}
            ~[tnpos=l]{\shortstack[r]{0.41\\\$1448}}
            \TC[fillcolor=yellow,fillstyle=solid]~{\makebox[0pt][c]{51}}
            ~[tnpos=r]{\shortstack[r]{0.76\\\$4365}}
          }
          \TC[fillcolor=green,fillstyle=solid]~{\makebox[0pt][c]{2037}}
          ~[tnpos=r]{\shortstack[r]{0.21\\\$1903}}
 }
 }
 }
 }
 \caption{GUIDE classification tree for predicting missingness in
   INTRDVX (INTRDVX\_ = C) from 4609 observations and 630
   $X$
   variables, with minimum node size 50.  At each split, an
   observation goes to the left branch if and only if the condition is
   satisfied.  The symbol `$\leq_*$'
   stands for `$\leq$
   or missing'.  Set $S_{1}$
   = \{CO, DE, FL, HI, IL, LA, MA, MO, NJ, NY, OH, PA, SC, TN, WA,
   WI\}; set $S_{2}$
   = \{1102, 1110, 1423\}.  PSU codes are given in
   Table~\ref{tab:psu}. 
Numbers beside each terminal node are the proportion missing
    INTRDVX (top) and the mean INTRDVX of the non-missing values; the
    sample size is beneath the node. Yellow and green nodes have
    proportions of INTRDVX missing greater and less, respectively,
    than 0.50.}
  \label{fig:GCT:630}
\end{figure}

\item[GCT.]  This uses a GUIDE classification tree \citep{guidec} to
  form adjustment cells for conditional mean
  imputation. Figure~\ref{fig:GCT:630} shows the GUIDE tree for
  predicting INTRDVX\_ = C or D.  The first split of the tree sends
  368 CUs with ERANKH either missing or $\leq
  0.006$ to the left node where 304 (83\%) of the CUs are missing
  INTRDVX (i.e., INTRDVX\_ = C).  The mean INTRDVX (\$952) among the 64
  CUs in the node with non-missing INTRDVX is used to impute the value
  of INTRDVX for the other 304 CUs in the node.  Repeating this at all
  the terminal nodes and substituting the imputed values in
  (\ref{eq:imp}) gives the estimate of $\mu$.
  The same answer can be obtained with the IPW method too.  Let
  $t$
  denote a terminal node of the tree and $p(t)$
  the proportion of non-missing $Y$
  in $t$.
  Then setting $\hat{\pi}
  = p(t)$ for all CUs in each
  $t$ in~(\ref{eq:ipw}) gives the same estimate.

\begin{figure}
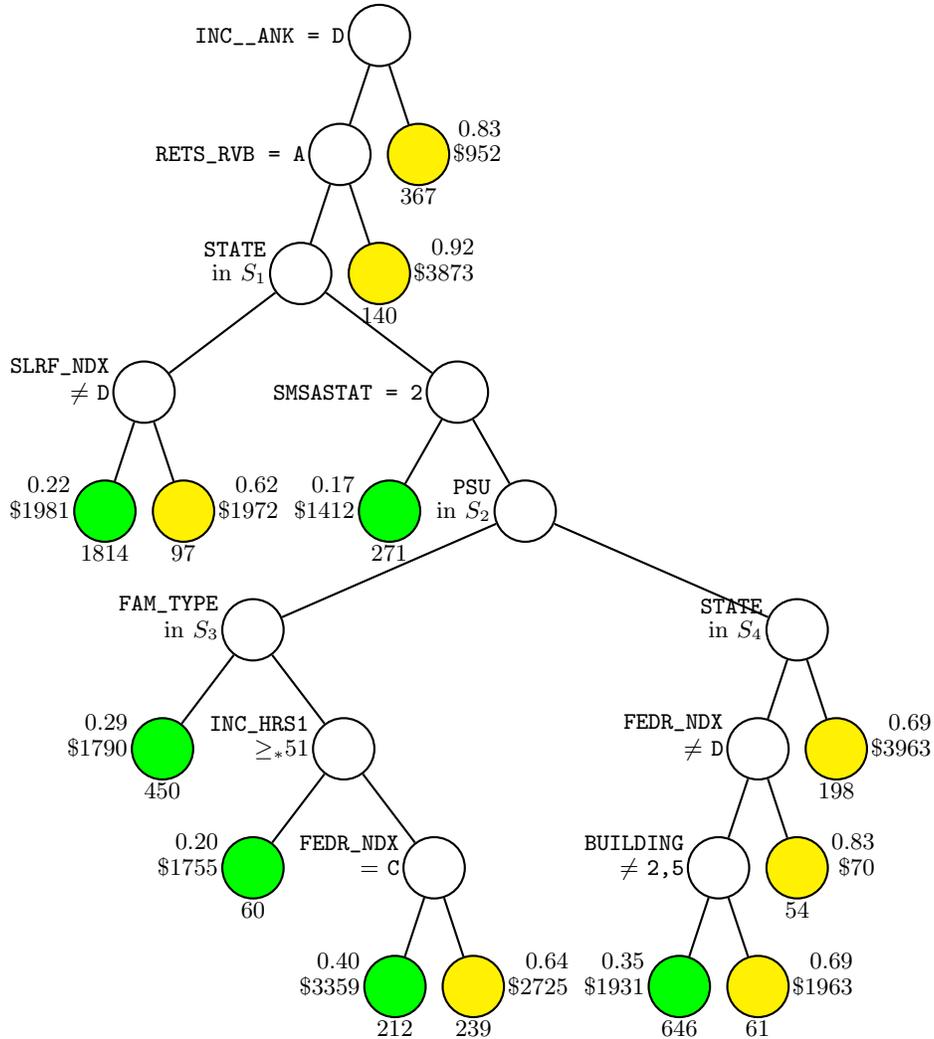

  \centering
  \psset{linecolor=black,tnsep=1pt,tndepth=0cm,tnheight=0cm,treesep=0.2cm,levelsep=45pt,radius=12pt}
  \pstree[treemode=D]{\TC~[tnpos=l]{\texttt{\detokenize{INC__ANK = D}}}}{
    \pstree[treemode=D]{\TC~[tnpos=l]{\texttt{\detokenize{RETS_RVB = A}}}}{
      \pstree[treemode=D]{\TC~[tnpos=l]{\shortstack[r]{\texttt{\detokenize{STATE}}\\in $S_{1}$}}}{
        \pstree[treemode=D]{\TC~[tnpos=l]{\shortstack[r]{\texttt{\detokenize{SLRF_NDX}}\\$\neq$ \texttt{D}}}}{
          \TC[fillcolor=green,fillstyle=solid]~{\makebox[0pt][c]{1814}}
          ~[tnpos=l]{\shortstack[r]{0.22\\\$1981}}
          \TC[fillcolor=yellow,fillstyle=solid]~{\makebox[0pt][c]{97}}
          ~[tnpos=r]{\shortstack[r]{0.62\\\$1972}}
        }
        \pstree[treemode=D]{\TC~[tnpos=l]{\texttt{\detokenize{SMSASTAT = 2}}}}{
          \TC[fillcolor=green,fillstyle=solid]~{\makebox[0pt][c]{271}}
          ~[tnpos=l]{\shortstack[r]{0.17\\\$1412}}
          \pstree[treemode=D]{\TC~[tnpos=l]{\shortstack[r]{\texttt{\detokenize{PSU}}\\in $S_{2}$}}}{
            \pstree[treemode=D]{\TC~[tnpos=l]{\shortstack[r]{\texttt{\detokenize{FAM_TYPE}}\\in $S_{3}$}}}{
              \TC[fillcolor=green,fillstyle=solid]~{\makebox[0pt][c]{450}}
              ~[tnpos=l]{\shortstack[r]{0.29\\\$1790}}
              \pstree[treemode=D]{\TC~[tnpos=l]{\shortstack[r]{\texttt{\detokenize{INC_HRS1}}\\$\geq_*$51}}}{
                \TC[fillcolor=green,fillstyle=solid]~{\makebox[0pt][c]{60}}
                ~[tnpos=l]{\shortstack[r]{0.20\\\$1755}}
                \pstree[treemode=D]{\TC~[tnpos=l]{\shortstack[r]{\texttt{\detokenize{FEDR_NDX}}\\= \texttt{\detokenize{C}}}}}{
                  \TC[fillcolor=green,fillstyle=solid]~{\makebox[0pt][c]{212}}
                  ~[tnpos=l]{\shortstack[r]{0.40\\\$3359}}
                  \TC[fillcolor=yellow,fillstyle=solid]~{\makebox[0pt][c]{239}}
                  ~[tnpos=r]{\shortstack[r]{0.64\\\$2725}}
                }
              }
            }
            \pstree[treemode=D]{\TC~[tnpos=l]{\shortstack[r]{\texttt{\detokenize{STATE}}\\in $S_{4}$}}}{
              \pstree[treemode=D]{\TC~[tnpos=l]{\shortstack[r]{\texttt{\detokenize{FEDR_NDX}}\\$\neq$ \texttt{D}}}}{
                \pstree[treemode=D]{\TC~[tnpos=l]{\shortstack[r]{\texttt{\detokenize{BUILDING}}\\$\neq$ \texttt{2,5}}}}{
                  \TC[fillcolor=green,fillstyle=solid]~{\makebox[0pt][c]{646}}
                  ~[tnpos=l]{\shortstack[r]{0.35\\\$1931}}
                  \TC[fillcolor=yellow,fillstyle=solid]~{\makebox[0pt][c]{61}}
                  ~[tnpos=r]{\shortstack[r]{0.69\\\$1963}}
                }
                \TC[fillcolor=yellow,fillstyle=solid]~{\makebox[0pt][c]{54}}
                ~[tnpos=r]{\shortstack[r]{0.83\\\$70}}
              }
              \TC[fillcolor=yellow,fillstyle=solid]~{\makebox[0pt][c]{198}}
              ~[tnpos=r]{\shortstack[r]{0.69\\\$3963}}
            }
          }
        }
      }
      \TC[fillcolor=yellow,fillstyle=solid]~{\makebox[0pt][c]{140}}      
      ~[tnpos=r]{\shortstack[r]{0.92\\\$3873}}
    }
    \TC[fillcolor=yellow,fillstyle=solid]~{\makebox[0pt][c]{367}}
    ~[tnpos=r]{\shortstack[r]{0.83\\\$952}}
  }
  \caption{RPART classification tree for predicting missingness in
    INTRDVX (INTRDVX\_ = C) from 4609 observations and 630 $X$
    variables, with minimum node size 50.  Set $S_{1}$ = \{AL, AK, AZ,
    CA, CT, DC, GA, ID, IN, KS, KY, ME, MD, MN, MO, NE, NV, NH, OR,
    TX, UT, VA, WV\}; $S_{2}$ = \{1103, 1111, 1207, 1208, 1210,
    1320\}; $S_3$ = \{2, 7, 8, 9\}; $S_4$ = \{AZ, DE, LA, NY, PA, SC,
    TN\}. Codes for PSU, FAM\_TYPE and BUILDING are given in
    Tables~\ref{tab:psu}, \ref{tab:famtype} and \ref{tab:building}.
    Numbers beside each terminal node are the proportion missing
    INTRDVX (top) and the mean INTRDVX of the non-missing values; the
    sample size is beneath the node. Yellow and green nodes have
    proportions of INTRDVX missing greater and less, respectively,
    than 0.50.}
  \label{fig:RCT:630}
\end{figure}

\item[RCT.] This is the RPART version of GCT where instead of GUIDE,
  RPART is used to construct the classification tree shown in
  Figure~\ref{fig:RCT:630}.  GUIDE differs from CART and RPART in
  several major respects, one being selection bias. Because CART
  employs greedy search to split each node, it is biased toward
  selecting variables that permit more splits of the data
  \citep{quest}. Examples are ordinal variables with many distinct
  values and categorical variables with many levels.  GUIDE does not
  have the bias. There is a hint of the bias if we compare the GCT and
  RCT trees. RCT splits on four categorical variables: STATE twice and
  PSU, FAM\_TYPE and BUILDING once each. GCT splits once on STATE and
  PSU and on no other categorical variables. Because STATE, PSU,
  FAM\_TYPE and BUILDING have 39, 21, 9, and 10 levels, respectively,
  they allow approximately $2^{38}$, $2^{20}$, $2^8$, and $2^9$ splits
  of the data. The codes for PSU, FAM\_TYPE and BUILDING are given in
  the Appendix. Another difference between GUIDE and CART is their
  treatment of missing values.  GUIDE sends all missing values on a
  split variable to one node or the other while CART and RPART use
  surrogate splits.  See \citet{guidec} for more information.

\item[GCF.]  This is an alternative IPW method to GCT where, instead
  of a single tree, a GUIDE classification forest \citep{ISI14} is
  used to estimate $\pi_i$.  GUIDE forest is an ensemble of 500
  unpruned classification trees, each constructed from a bootstrap
  sample of the data to predict whether INTRDVX is missing.  The
  Random forest \citep{randomforest01} R implementation \citep{RF}
  cannot be used directly here because: (1)~it requires missing $X$
  values to be imputed beforehand and (2)~it does not allow
  categorical variables with more than 32 levels.

\item[GRT.] This is a conditional mean imputation method that uses a
  GUIDE piecewise-constant regression tree \citep{guide} to impute
  missing INTRDVX values. Unlike GCT and GCF, it uses only the subset
  of 2838 CUs with non-missing INTRDVX. The tree is shown in
  Figure~\ref{fig:GRT:630}.  It splits first on AGE\_REF = missing or
  $\leq$ 57. About half of the 2838 CUs satisfy this condition; their
  mean (\$907) is used to impute the missing INTRDVX values in the
  node. Repeating this procedure at each terminal node yields a
  completed set of INTRDVX values for application in~(\ref{eq:imp}).

\begin{figure}
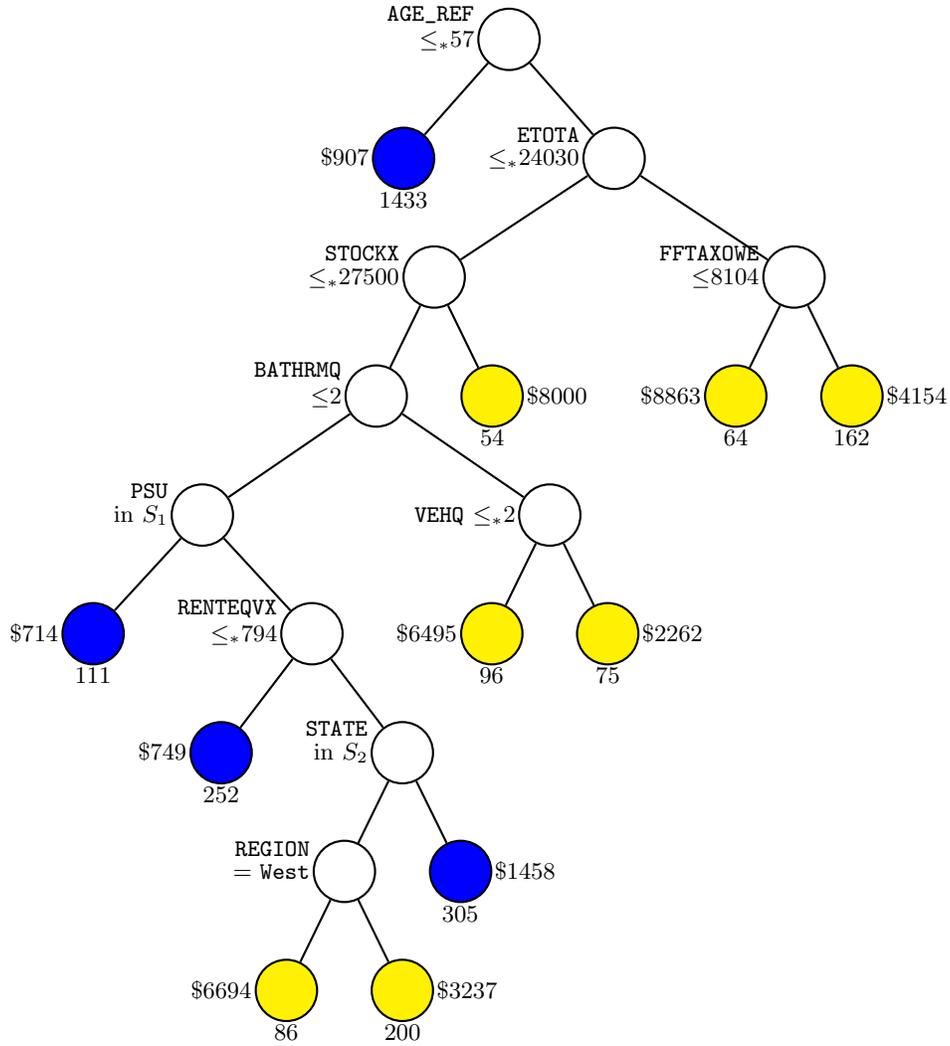

  \centering
  \psset{linecolor=black,tnsep=1pt,tndepth=0cm,tnheight=0cm,treesep=0.7cm,levelsep=45pt,radius=12pt}
  \pstree[treemode=D]{\TC~[tnpos=l]{\shortstack[r]{\texttt{\detokenize{AGE_REF}}\\$\leq_*$57}}}{
    \TC[fillcolor=blue,fillstyle=solid]~{\makebox[0pt][c]{1433}}~[tnpos=l]{\$907}
    \pstree[treemode=D]{\TC~[tnpos=l]{\shortstack[r]{\texttt{\detokenize{ETOTA}}\\$\leq_*$24030}}}{
      \pstree[treemode=D]{\TC~[tnpos=l]{\shortstack[r]{\texttt{\detokenize{STOCKX}}\\$\leq_*$27500}}}{
        \pstree[treemode=D]{\TC~[tnpos=l]{\shortstack[r]{\texttt{\detokenize{BATHRMQ}}\\$\leq$2}}}{
          \pstree[treemode=D]{\TC~[tnpos=l]{\shortstack[r]{\texttt{\detokenize{PSU}}\\in $S_{1}$}}}{
            \TC[fillcolor=blue,fillstyle=solid]
            ~{\makebox[0pt][c]{111}}~[tnpos=l]{\$714}
            \pstree[treemode=D]{\TC~[tnpos=l]{\shortstack[r]{\texttt{\detokenize{RENTEQVX}}\\$\leq_*$794}}}{
              \TC[fillcolor=blue,fillstyle=solid]
              ~{\makebox[0pt][c]{252}}~[tnpos=l]{\$749}
              \pstree[treemode=D]{\TC~[tnpos=l]{\shortstack[r]{\texttt{\detokenize{STATE}}\\in $S_{2}$}}}{
                \pstree[treemode=D]{\TC~[tnpos=l]{\shortstack[r]{\texttt{\detokenize{REGION}}\\= \texttt{\detokenize{West}}}}}{
                  \TC[fillcolor=yellow,fillstyle=solid]
                  ~{\makebox[0pt][c]{86}}~[tnpos=l]{\$6694}
                  \TC[fillcolor=yellow,fillstyle=solid]
                  ~{\makebox[0pt][c]{200}}~[tnpos=r]{\$3237}
                }
                \TC[fillcolor=blue,fillstyle=solid]
                ~{\makebox[0pt][c]{305}}~[tnpos=r]{\$1458}
              }
            }
          }
          \pstree[treemode=D]{\TC~[tnpos=l]{\texttt{\detokenize{VEHQ}} $\leq_*$2}}{
            \TC[fillcolor=yellow,fillstyle=solid]
            ~{\makebox[0pt][c]{96}}~[tnpos=l]{\$6495}
            \TC[fillcolor=yellow,fillstyle=solid]
            ~{\makebox[0pt][c]{75}}~[tnpos=r]{\$2262}
          }
        }
        \TC[fillcolor=yellow,fillstyle=solid]
        ~{\makebox[0pt][c]{54}}~[tnpos=r]{\$8000}
      }
      \pstree[treemode=D]{\TC~[tnpos=l]{\shortstack[r]{\texttt{\detokenize{FFTAXOWE}}\\$\leq$8104}}}{
        \TC[fillcolor=yellow,fillstyle=solid]
        ~{\makebox[0pt][c]{64}}~[tnpos=l]{\$8863}
        \TC[fillcolor=yellow,fillstyle=solid]
        ~{\makebox[0pt][c]{162}}~[tnpos=r]{\$4154}
      }
    }
  }
  \caption{GUIDE regression tree for predicting INTRDVX from 2838
    observations and 630 $X$ variables, with minimum node size 50.  At
    each split, an observation goes to the left branch if and only if
    the condition is satisfied.  The symbol `$\leq_*$' stands for
    `$\leq$ or missing'.  Set $S_{1}$ = \{1102, 1103, 1109, 1313,
    1419, 1420, 1429\}; $S_{2}$ = \{AL, AZ, CA, CT, FL, GA, ID, IN,
    KY, ME, MD, MI, MN, MO, NV, OH, OR, SC, TX, WA, WI\}.  Codes for
    PSU are given in Table~\ref{tab:psu}.  Sample sizes and means of
    INTRDVX are printed below and beside nodes. Blue and yellow nodes
    have mean INTRDVX below and above, respectively, the SIM estimate
    of \$2009.}
  \label{fig:GRT:630}
\end{figure}

\item[RRT.] This is the RPART version of GRT where RPART is used
  instead of GUIDE. The tree, shown in Figure~\ref{fig:RRT:630}, has
  the same top split as the GRT tree. It subsequently splits on STATE
  three times, likely because STATE permits more splits.

\begin{figure}
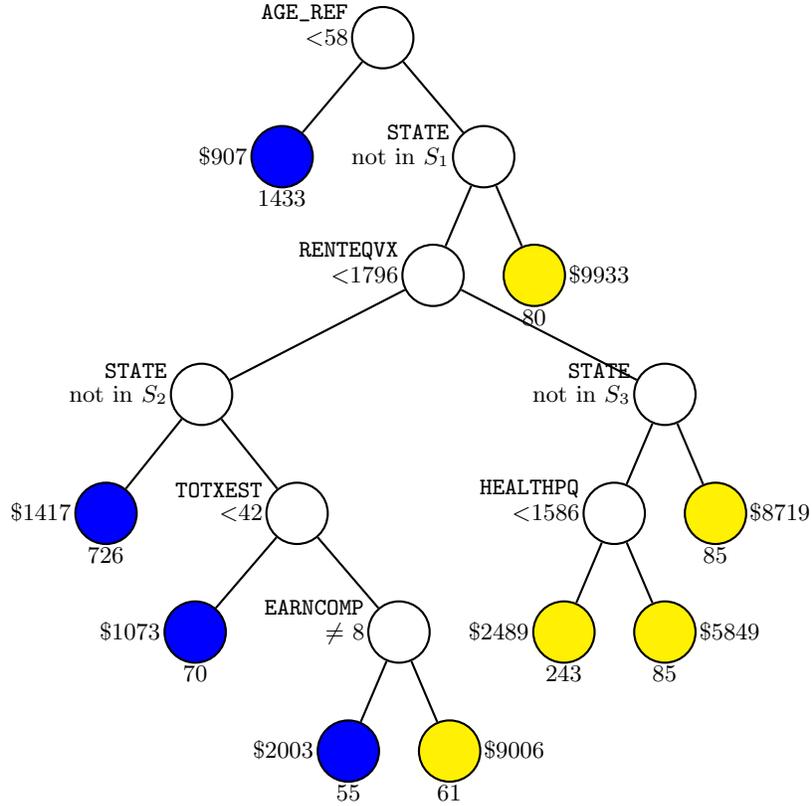

  \centering
  \psset{linecolor=black,tnsep=1pt,tndepth=0cm,tnheight=0cm,treesep=0.5cm,levelsep=45pt,radius=12pt}
  \pstree[treemode=D]{\TC~[tnpos=l]{\shortstack[r]{\texttt{\detokenize{AGE_REF}}\\$<$58}}}{
    \TC[fillcolor=blue,fillstyle=solid]~{\makebox[0pt][c]{1433}}~[tnpos=l]{\$907}
    \pstree[treemode=D]{\TC~[tnpos=l]{\shortstack[r]{\texttt{\detokenize{STATE}}\\not in $S_{1}$}}}{
      \pstree[treemode=D]{\TC~[tnpos=l]{\shortstack[r]{\texttt{\detokenize{RENTEQVX}}\\$<$1796}}}{
        \pstree[treemode=D]{\TC~[tnpos=l]{\shortstack[r]{\texttt{\detokenize{STATE}}\\not in $S_{2}$}}}{
          \TC[fillcolor=blue,fillstyle=solid]~{\makebox[0pt][c]{726}}~[tnpos=l]{\$1417}
          \pstree[treemode=D]{\TC~[tnpos=l]{\shortstack[r]{\texttt{\detokenize{TOTXEST}}\\$<$42}}}{
            \TC[fillcolor=blue,fillstyle=solid]~{\makebox[0pt][c]{70}}~[tnpos=l]{\$1073}
            \pstree[treemode=D]{\TC~[tnpos=l]{\shortstack[r]{\texttt{\detokenize{EARNCOMP}}\\$\neq$ 8}}}{
              \TC[fillcolor=blue,fillstyle=solid]~{\makebox[0pt][c]{55}}~[tnpos=l]{\$2003}
              \TC[fillcolor=yellow,fillstyle=solid]~{\makebox[0pt][c]{61}}~[tnpos=r]{\$9006}
            }
          }
        }
        \pstree[treemode=D]{\TC~[tnpos=l]{\shortstack[r]{\texttt{\detokenize{STATE}}\\not in $S_{3}$}}}{
          \pstree[treemode=D]{\TC~[tnpos=l]{\shortstack[r]{\texttt{\detokenize{HEALTHPQ}}\\$<$1586}}}{
            \TC[fillcolor=yellow,fillstyle=solid]~{\makebox[0pt][c]{243}}~[tnpos=l]{\$2489}
            \TC[fillcolor=yellow,fillstyle=solid]~{\makebox[0pt][c]{85}}~[tnpos=r]{\$5849}
          }
          \TC[fillcolor=yellow,fillstyle=solid]~{\makebox[0pt][c]{85}}~[tnpos=r]{\$8719}
        }
      }
      \TC[fillcolor=yellow,fillstyle=solid]~{\makebox[0pt][c]{80}}~[tnpos=r]{\$9933}
    }
  }
  \caption{RPART regression tree for predicting INTRDVX from from 2838
    observations and 630 $X$
    variables, with minimum node size 50.  At each split, an
    observation goes to the left branch if and only if the condition
    is satisfied.  Set $S_{1}$
    = \{CO, ME, NV, SC, WA, WV\}; $S_{2}$
    = \{FL, GA, MO, NH, OH, OR, VA, WI\}; $S_3$
    = \{AL, CT, ID, IL, KY, MO, NY, TX\}.  Codes for EARNCOMP are in
    Table~\ref{tab:earncomp}.  Sample sizes and means of INTRDVX are
    printed below and beside nodes. Blue and yellow nodes have mean
    INTRDVX below and above, respectively, the SIM estimate of
    \$2009.}
  \label{fig:RRT:630}
\end{figure}

\item[GRF.]  This is an alternative to GRT. It uses a GUIDE regression
  forest \citep{lchen12,ISI14} of 500 unpruned regression trees to
  carry out conditional mean imputation.

\item[GMICE.]  CART-MICE cannot be used for the CE data due to
  limitations in the CART algorithm and software.  The R TREE package
  \citep{tree} used in CART-MICE cannot handle categorical variables
  with more than 32 levels.  RPART \citep{rpart} does not have this
  limitation, but both are too computationally expensive for
  imputation of categorical variables with many categories.  GMICE is
  CART-MICE with GUIDE in place of CART, with ten iterations for each
  imputation.
\end{description}

To understand the differences between CART-MICE and GMICE, recall that
at each node, CART searches for the best split on every $X$ variable
and then chooses the split that most reduces node impurity.  Suppose
$Y$ is categorical with more than two categories and $X$ is also
categorical with $p$ categories. CART has to search through
$(2^{p-1}-1)$ splits of the form ``$X \in A$'' to find the subset $A$
that yields the best split on $X$.  Because STATE has 39 categories in
our data, there are $2^{38}-1 \approx 2 \times 10^{11}$ splits.  Other
variables with large numbers of categories are HHID (household
identifier), PSU, OCCUCOD1, and OCCUCOD2 (spouse occupation), with 46,
21, 15, and 15 levels, respectively.

GUIDE gets around the computation problem with a two-step approximate
solution.  In the first step, GUIDE selects an $X$ variable to split
the node by means of chi-squared tests of association with $Y$. In the
second step, GUIDE finds a split on the selected $X$. Therefore GUIDE
does \emph{not} search for the best split on $X_i$ for every
$X_i$. There are two important benefits to this approach. One is the
obvious significant reduction in computation time. The other, which is
more subtle, is the first step eliminates the selection bias inherent
in the CART approach \citep{quest,cruise}.

Suppose in the second step that the selected $X$ variable has $p$
categories $\{c_1, c_2, \ldots, c_p\}$ and $Y$ has $q$ categories
$\{j_1, j_2, \ldots, j_q\}$.  GUIDE uses the following sequential
procedure that reduces the search space if $p$ is large and $q > 2$.
\begin{enumerate}
\item If $q=2$, let $c_1', c_2', \ldots, c_p'$ be the ordered
  categories of $X$ such that
  $\#(Y=j_1 , X=c_1') \leq \#(Y=j_1 , X=c_2') \leq \ldots \leq
  \#(Y=j_1 , X=c_p')$
  and $A_i = \{c_1', c_2', \ldots, c_i'\}$, $i=1,2,\ldots,p$.  Search
  for the best split among the $(p-1)$ splits $\{X \in A_i\}$
  \citep[p.~101]{cart}.
\item Otherwise, if $p \leq 11$, search through all
  $(2^{p-1}-1) \leq 1023$ splits on $X$.
\item Otherwise, if $2 < q \leq 11$ and $p > 20$, define a new
  categorical variable $X'$ by merging the categories of $X$ into $q$
  categories. Specifically, for $k = 1, 2, \ldots, p$, let $j_k'$ be
  the predicted value of $Y$ that minimizes the node impurity among
  the observations with $X=c_k$ in the node. Define
  $X' = \sum_k j_k' I(X=c_k)$ and split the node on $X'$, which has
  $2^{q-1}-1 \leq 1023$ splits.
\item If none of the above conditions holds, use linear discriminant
  analysis on $Y$ to transform $X$ into an ordered variable
  $X''$. Define indicator variables $D_k = I(X = c_k)$,
  $k=1,2,\ldots, p$, and $X'' = \sum_k a_k D_k$, where
  $(a_1, a_2, \ldots, a_p)$ is the normalized vector corresponding to
  the largest discriminant coordinate (equivalent to maximizing the
  ANOVA F-statistic of $Y$ on $X''$).  Find the split on $X''$ that
  most reduces node impurity. Because $X''$ takes at most $p$ ordered
  values, the search is on only $(p-1)$ splits. Further, because $X''$
  induces an order on the unordered $X$ values, a split of the form
  $X'' \leq b$ can be expressed in the form $X \in A$. This technique
  is borrowed from \citet{lv88}.
\end{enumerate}

Another important difference between GUIDE and CART is how they deal
with missing values.  If $X$ has missing values, GUIDE creates a
``missing'' level to use in the chi-squared tests for variable
selection as well as for split set selection. As a result, all
observations are used.  CART uses surrogates splits, which has been
shown to induce a selection bias on variables that have more missing
values \citep{cruise}. Besides, there is empirical evidence that the
GUIDE approach to missing values yields higher average classification
accuracy than the RPART implementation of CART \citep{guidec}.

To accommodate MICE which works only for 19 variables in
Table~\ref{tab:19var}, we applied the methods to three nested sets of
$X$ variables.  The smallest is this set of 19. It is far from an
ideal test-bed because there are only 5 variables with missing values
(and 4 of them have trivially small numbers of missing values).  The
second set consists of 52 $X$ variables, obtained by combining three
groups of variables: the 19 in Table~\ref{tab:19var} and the top 20
$X$ variables determined by the GUIDE importance ranking method
\citep{lchen12,lhm15} for predicting INTRDVX\_ and INTRDVX,
respectively; see Tables~\ref{tab:top:class} and
\ref{tab:top:reg}. The third set is the full set of 630 variables.

\begin{table}
  \centering
  \caption{Estimates of mean INTRDVX (in dollars) and computation
    times for 3 nested sets of predictor variables. Estimated mean for SIM is \$2009. 
    Columns are ordered by computation time.  Results for AMELIA and MICE are based on
    5 multiple imputations. For 630 variables, AMELIA did not produce any imputations
     after 33 days.}
  \vspace{0.5em}
  \begin{tabular}{lrrrrrrrrr}
    & RRT & RCT & GRT & GCT & GCF & GRF & GMICE & AME & MICE \\ \hline
    \multicolumn{10}{c}{19 predictor variables} \\
    Mean & 2020 & 2009 & 2087 & 2151 & 2002 & 2109 & 2089 & 2122 & 2201 \\
    Time & 0.2s & 0.1s & 3s & 3s & 9s & 207s & 54s & 147s & 443s \\ \hline
    \multicolumn{10}{c}{52 predictor variables} \\
    Mean & 2027 & 2069 & 1941 & 2112 & 1897 & 2091 & 2132 & 2009 & Fail \\
    Time & 0.5s & 2s & 6s & 8s & 18s & 307s & 429s & 12h & Fail \\ \hline
    \multicolumn{10}{c}{630 predictor variables} \\ 
    Mean & 2013 & 2059 & 2058 & 2100 & 1886 & 2038 & 2032 & ? & Fail \\
    Time & 5s & 20s & 118s & 134s & 259s & 29m & 27h & 33d+ & Fail \\ \hline
  \end{tabular}
  \label{tab:popmeans}
\end{table}

\begin{figure}
  \centering
  \resizebox{\textwidth}{!}{\includegraphics*[94,276][505,520]{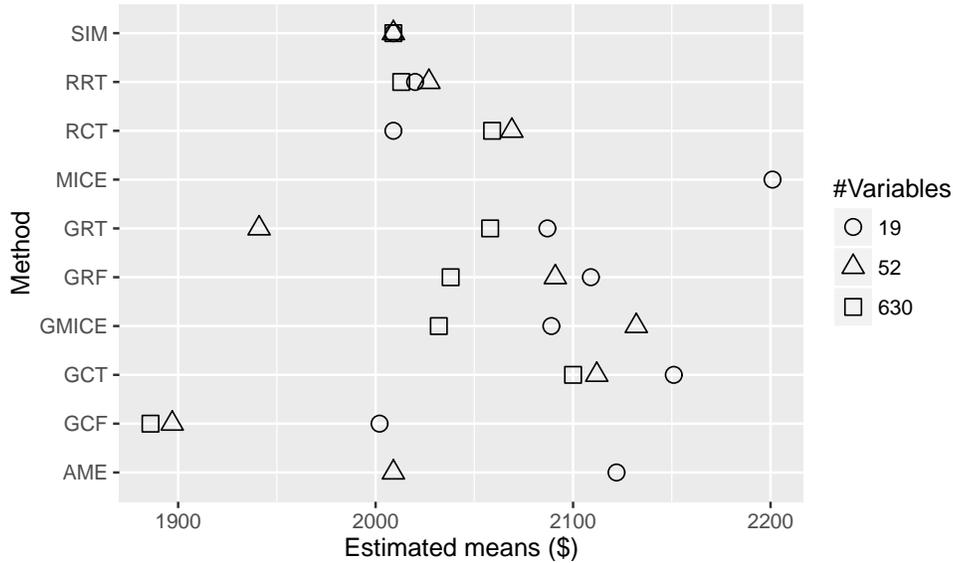}}
  \caption{Estimated mean INTRDVX of ten methods for CE data using 19,
    52 and 630 $X$ variables}
  \label{fig:allmeans}
\end{figure}

SIM estimates the mean INTRDVX as \$2009 for all three sets because it
does not use $X$ variables. Results for the other methods are shown in
Table~\ref{tab:popmeans} and graphed in Figure~\ref{fig:allmeans}.
Every method works on the set of 19 variables but MICE fails for the
other two sets. The mean estimates range from a low of \$1886 for GCF
to a high of \$2201 for MICE.  Computation times (measured on a Linux
computer with a 2.4 GHz AMD Opteron 16-core processor and 64 GB
memory) vary greatly.  MICE is the slowest on the only set of
variables for which it does not fail. It takes 443 sec.\ for five
imputations---more than 4000 times slower than RCT, which is the
fastest at 0.1 sec. AMELIA is the next slowest at 147 sec.\ for this
set. For the other two larger sets, AMELIA is the slowest, taking 12
hours for 52 variables and five imputations. For 630 variables, AMELIA
ran for 33 days without producing a single imputed data set. In the
next section, we report on a series of simulation experiments to
compare the bias and accuracy of the estimates.

\section{Simulation study}

\subsection{Experimental design}
\label{sec:setup}
Simulation studies whose goal is to evaluate methods under realistic
conditions usually start with a real data set $\mathcal{D}$ and then
generate artificial populations and samples from it in two steps:
\begin{description}
\item[Step I.]  Impute the missing $Y$ values in $\mathcal{D}$ and
  treat the resulting data set as a finite population $\mathcal{P}_1$
  from which the mean $\mu$ of $Y$ is computed.
\item[Step II.] Generate a population $\mathcal{P}_2$ from
  $\mathcal{P}_1$ by making some $X$ and $Y$ values in $\mathcal{P}_1$
  missing.
\end{description}
If the aim is for $\mathcal{P}_2$ to be as similar to $\mathcal{D}$ as
possible, the choice of methods in these two steps is critical.
Typically, MICE is used in Step I and a logistic regression propensity
model is fitted to the missing value flag variable in $\mathcal{D}$ to
estimate the probability that the variable is missing in Step~II.
These choices have the three undesirable consequences.
\begin{enumerate}
\item Using MICE in Step~I limits the number of variables with missing
  values to no more than a few, due to problems with
  multi-collinearity and quasi-complete separation. This makes it
  impossible to apply to data sets such as ours, without preselection
  of variables.  Besides, to impute a $Y$ variable, MICE necessarily
  imputes all $X$ variables with linear and logistic regression
  models. This forces relationships among the $X$ variables in
  $\mathcal{P}_1$ that do not exist in $\mathcal{D}$, thus producing
  simulated data that look more unlike the real data.
\item In Step~II, logistic regression propensity models cannot be
  constructed from $X$ variables with missing values.  The latter must
  be imputed first or the propensity models must be built from subsets
  of variables or subsets of data. Neither solution is desirable.
  Imputing the $X$ variables (e.g., with MICE) distorts the data and
  building propensity models from subsets of data requires the
  artificial assumption that the $X$ variables are MAR. Some studies
  solve this problem by using only completely observed $X$ variables
  for propensity modeling \citep{burgette10}, but this is artificial
  too because the probability of a missing value in $Y$ often depends
  on $X$ variables with missing values.  For example, whether or not
  interest and dividends is missing depends on the values and
  missingness of variables such as salary, income taxes paid, and
  value of stocks.
\item Standard linear and logistic regression models with prespecified
  fixed sets of predictor variables are inapplicable if the number of
  variables exceeds the sample size.
\end{enumerate}

The difficulties vanish if GUIDE forests are used instead of MICE and
logistic regression in Steps~I and II respectively.
\begin{description}
\item[Step I.]  Fit a GUIDE \emph{regression} forest to the 2838 CUs
  in $\mathcal{D}$ with non-missing INTRDVX values and let $\sigma^2$
  be the mean squared residual of the fitted model. Let $\hat{y}_i$
  ($i=1,2,\ldots, 1771$) denote the predicted value of the $i$th CU in
  $\mathcal{D}$ with missing INTRDVX and $\epsilon_i$ be an
  independent random number drawn from a normal distribution with mean
  0 and variance $\sigma^2$.  Impute the missing INTRDVX value with
  $\max(\hat{y}_i + \epsilon_i, 0)$.  Adding $\epsilon_i$ to
  $\hat{y}_i$ prevents the imputed values from being more smooth than
  the non-missing values. It also produces a different $\mathcal{P}_1$
  for each simulation trial.  Truncation at 0 ensures that all INTRDVX
  values are nonnegative.  Non-missing INTRDVX values in $\mathcal{D}$
  are carried over unchanged to $\mathcal{P}_1$.
\item[Step II.]  Fit a GUIDE \emph{classification} forest to all 4609
  members of $\mathcal{D}$, using the missing value flag INTRDVX\_ as
  dependent variable. Use the predicted values from the model to
  estimate the probability that INTRDVX is missing for each member of
  $\mathcal{D}$. Finally, make the value of INTRDVX (original or
  imputed) MAR in $\mathcal{P}_2$ according to these probabilities.
\end{description}
A principal advantage of GUIDE forests over other methods, including
Random forest, is that no imputation is required of missing $X$
values. As a result, their missing mechanisms in the simulated data
are the same as in the real data, i.e., not MAR.  Other advantages are
fast computation speed, unlimited by number of variables and sample
size, and absence of the multi-collinearity and quasi-complete
separation difficulties that afflict linear and logistic regression.

In our simulation experiments, $\mathcal{D}$ is the CE data set of
4609 CUs with 19, 52, or 630 $X$ variables.  A GUIDE regression forest
is fitted to each choice of $\mathcal{D}$ and a population
$\mathcal{P}_1$ obtained.  Each simulation trial then consists of
generating $\mathcal{P}_2$ and drawing a simple random sample without
replacement from it. We use sampling fractions of 5\%, 10\%, and 25\%
(corresponding to sample sizes 230, 461, and 1152).  Finally, each
method being evaluated is applied to the sample to impute the missing
INTRDVX values (for conditional mean imputation) or estimate $\pi_i$
(for IPW) and an estimate $\hat{\mu}$ of $\mu$ computed.

\subsection{Results}
\label{sec:results}
\begin{figure}
  \centering
  \resizebox{\textwidth}{!}{\includegraphics*[19,35][575,752]{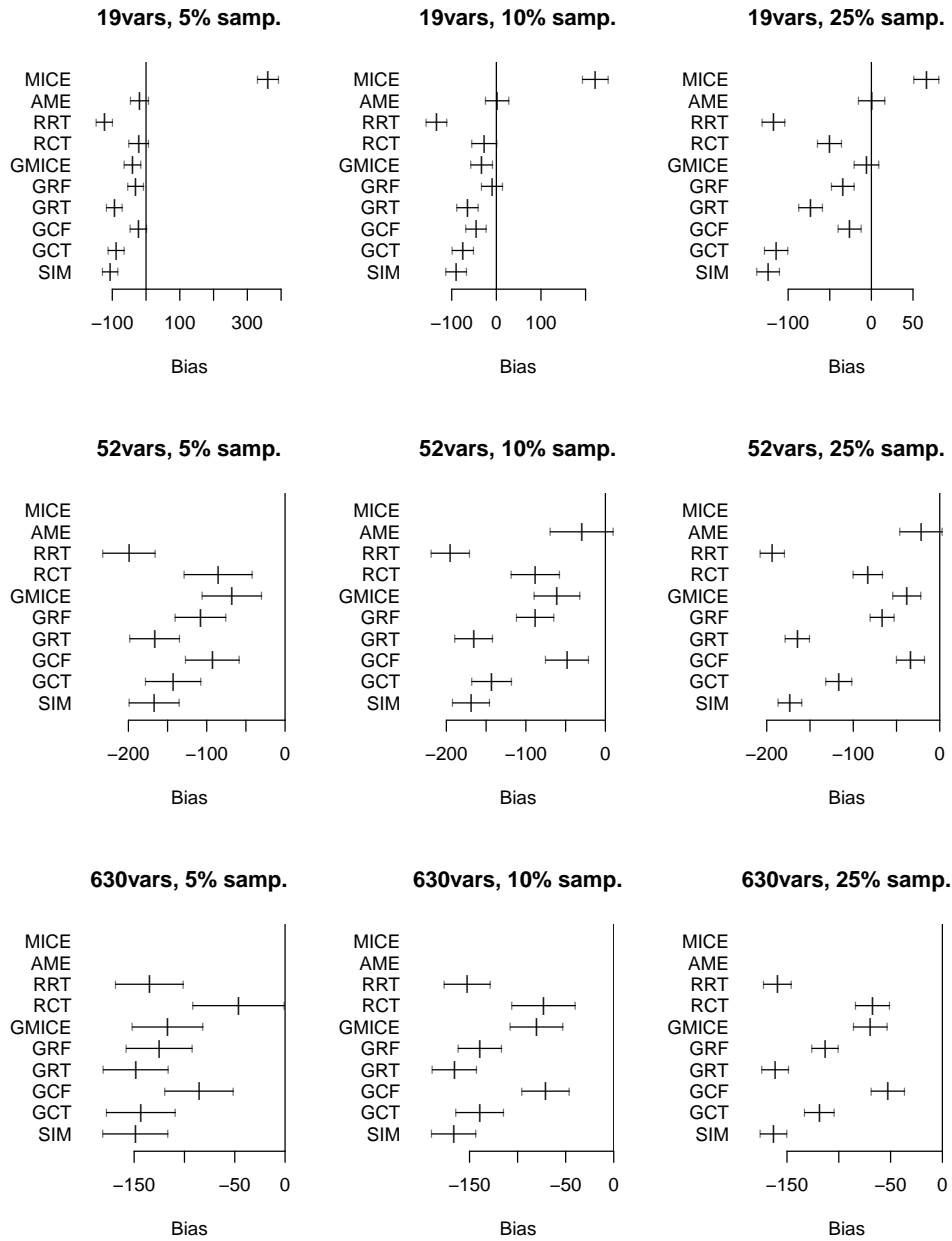}}
  \caption{Bias (with 95\% confidence intervals) for populations
    generated from 19, 52 and 630 $X$ variables.  MICE fails for 52
    and 630 variables; AMELIA fails for 630 variables and for 52
    variables at 5\% sampling.}
  \label{fig:biasnonnull}
\end{figure}

Let $M$ denote the number of simulation trials, where $M=500$ except
for the case with 19 $X$ variables and 5\% sampling when $M=1000$.
For trial $m$, let $\mu_m$ denote the mean of $\mathcal{P}_1$ (which
varies with $m$) and let its estimate be $\hat{\mu}_m$ for a given
method. The bias and root mean squared error (RMSE) of the method are
estimated by
\begin{eqnarray*}
  \mbox{Bias} & = & M^{-1} \sum_m (\hat{\mu}_m-\mu_m) \\
  \mbox{RMSE} & = & \left\{M^{-1} \sum_m (\hat{\mu}_m-\mu_m)^2\right\}^{1/2}.
\end{eqnarray*}
Figure~\ref{fig:biasnonnull} plots the estimated biases with 95\%
confidence intervals. We observe that:
\begin{enumerate}
\item AMELIA has the smallest bias in the situations where it works,
  viz., where sample size is large relative to number of variables.
\item MICE is the only method having positive bias. In the only
  situation where it works (19 $X$ variables), MICE has the largest
  bias at 5\% and 10\% sampling.  This is surprising given that there
  is only one $X$ variable with a significant amount of missing values
  here.
\item GMICE is always better than MICE in terms of bias. The former is
  consistently among the top three methods.
\item RRT has the next largest bias after MICE. Its bias can be larger
  than that of SIM, which does not use $X$ variables.  On the other
  hand, RCT has relatively low bias.
\item Comparing RPART versus GUIDE regression trees, RRT has larger
  bias than GRT for 19 and 52 variables; the two are about the same
  for 630 variables. The reverse is true for classification trees,
  with RCT doing consistently better than GCT.
\item Among the GUIDE methods, GCF tends to have the smaller bias than
  GCT, GRT, and GRF.
\end{enumerate}

\begin{figure}
  \centering
  \resizebox{\textwidth}{!}{\includegraphics*[18,34][592,762]{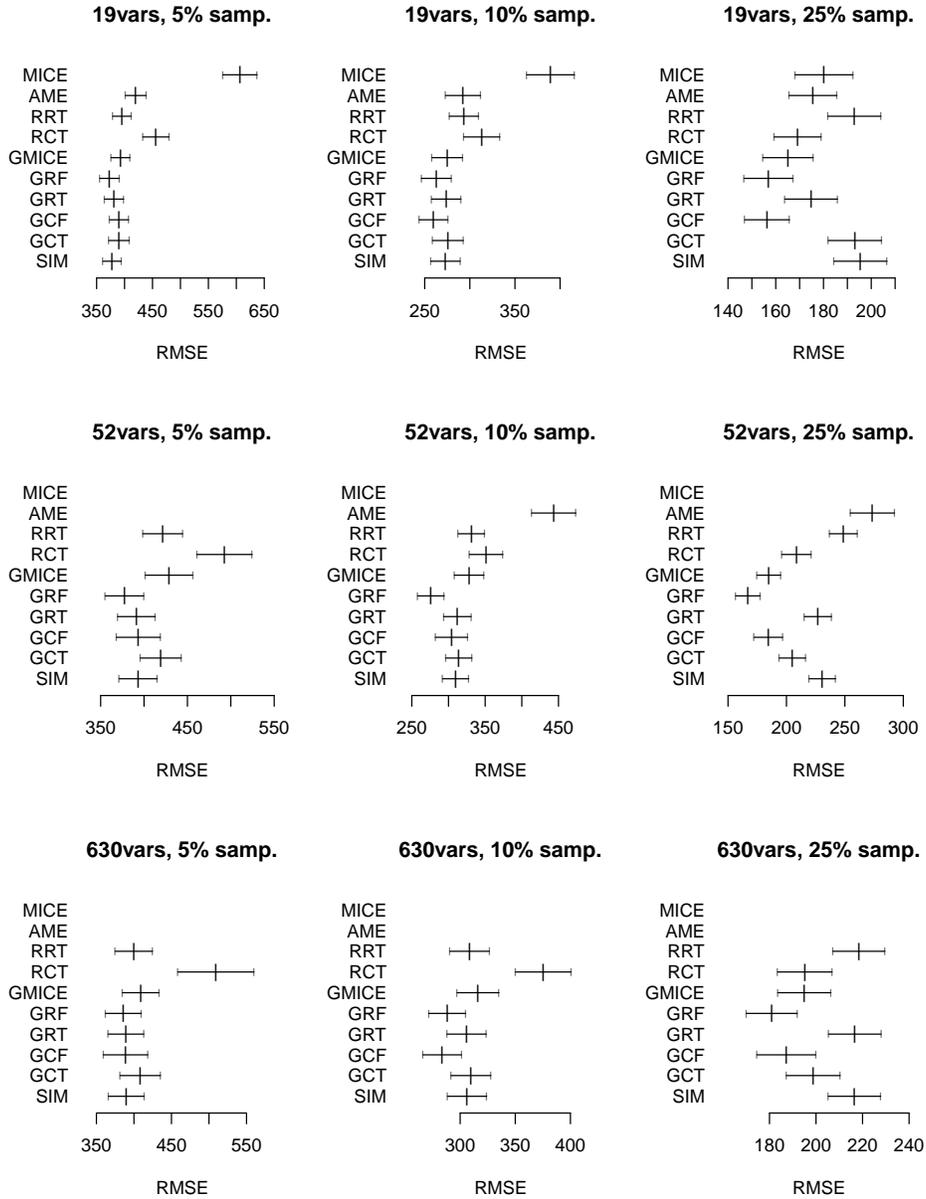}}
  \caption{RMSE of methods (with 95\% confidence intervals) using
    populations generated from 19, 52 and 630 $X$ variables.  MICE
    fails for 52 and 630 variables; AMELIA fails for 630 variables and
    for 52 variables at 5\% sampling.}
  \label{fig:rmsenonnull}
\end{figure}

The RMSE results shown in Figure~\ref{fig:rmsenonnull} reveal another
aspect of the methods:
\begin{enumerate}
\item MICE is again worst in two of the three situations where it
  works (5\% and 10\% sampling with 19 variables).
\item AMELIA is not as good in RMSE as it is in bias---its RMSE is
  largest for 52 variables and 10\% and 25\% sampling (it fails due to
  inadequate sample size for 5\% sampling).
\item RCT and RRT tend to have larger RMSEs.
\item The best method in terms of RMSE is GRF. It is followed by GCF
  and GMICE.
\end{enumerate}

Overall, GCF and GRF have the best combination of bias and RMSE, and
GMICE is third. When $\mathcal{D}$ has 19 variables, MICE has the
largest bias and RMSE for 5\% and 10\% sampling, though they are much
improved for 25\%. Its failure to work for larger numbers of variables,
however, is its biggest weakness. AMELIA has the lowest bias but the
highest RMSE when there are 52 variables.  

Table~\ref{tab:cpu} shows the average computational time to perform
one simulation trial for each method. If $\mathcal{D}$ has 19
variables, MICE is the slowest, taking 3 min.\ per simulation trial
for 1152 observations (25\% sampling). This is 4.4 times longer than
AMELIA, the next slowest. GRF is just behind AMELIA in third place;
GMICE is fourth, being 9.5 times faster than MICE.  If $\mathcal{D}$
has 52 variables (where MICE is out of contention), the difference in
average time per simulation trial for sample size 1152 is 2.6 hours
for AMELIA and 2.6 min.\ for GMICE.  Finally, if $\mathcal{D}$ has 630
variables, MICE and AMELIA are both out of contention. Then GMICE is
slowest at 62 hours per simulation trial and the corresponding times
for GCF and GRF, the next two slowest, are 33.7 and 32.3 sec.,
respectively. The fastest methods are RCT and RRT, which take from
fractions of a second to at most 4 sec.

\begin{table}[htbp]
  \centering
  \caption{Average computation times (sec.) per simulation trial}
  \label{tab:cpu} \vspace{0.5em}
  \begin{tabular}{rr|rrrrrrrrr}
    Frac & $n$ & RCT & RRT & GCT & GCF & GRT & GRF & GMICE & AME & MICE \\ \hline
    & & \multicolumn{9}{c}{19 variables} \\
    5\% & 230 & 0.02 & 0.02 & 0.5 & 1.6 & 0.2 & 2.1 & 3.2 & 7.3 & 70 \\
    10\% & 461 & 0.05 & 0.02 & 1.1 & 3.9 & 0.4 & 6.3 & 7.3 & 16.3 & 113 \\
    25\% & 1152 & 0.10 & 0.05 & 3.4 & 19.0 & 1.9 & 31.4 & 19.4 & 41.1 & 181 \\ \hline
    & & \multicolumn{9}{c}{52 variables} \\
    5\% & 230 & 0.05 & 0.03 & 3.2 & 3.3 & 1.0 & 4.2 & 31.1 & FAIL & FAIL \\
    10\% & 461 & 0.11 & 0.05 & 5.0 & 6.2 & 1.5 & 9.5 & 65.6 & 4272 & FAIL \\
    25\% & 1152 & 0.32 & 0.11 & 10.2 & 22.3 & 4.1 & 37.1 & 156.9 & 9323 & FAIL \\ \hline
    & & \multicolumn{9}{c}{630 variables} \\
    5\% & 230 & 0.57 & 0.33 & 6.0 & 33.3 & 5.8 & 27.8 & 2810 & FAIL & FAIL \\
    10\% & 461 & 1.27 & 0.56 & 12.6 & 94.2 & 12.0 & 77.7 & 6957 & FAIL & FAIL \\
    25\% & 1152 & 3.97 & 1.45 & 33.7 & 319.7 & 32.3 & 278.1 & 22204 & FAIL & FAIL \\ \hline
  \end{tabular}
\end{table}

\section{Conclusion}
\label{sec:conclusion}
We introduced several techniques to use classification and regression
methods for mean estimation of incomplete data. Some employ regression
trees to estimate conditional means in adjustment cells defined by the
nodes of the trees.  Others employ classification trees to estimate
missing propensities for inverse probability weighting.  Using a data
set from the U.S. Consumer Expenditure Survey as a test bed, we
performed a large simulation experiment to compare the methods with
AMELIA and MICE. A major goal in the experimental design is the
novelty of ensuring that the predictor variables are naturally
missing, i.e., not constrained to be MAR, in the simulation
populations. This is achieved by using GUIDE forests, which leave the
predictor-value missingness patterns intact.

The results demonstrate that classification and regression tree
methods have the following desirable properties that make them
deserving of serious consideration for analysis of incomplete data.
\begin{enumerate}
\item They are often superior to traditional methods (as represented
  by AMELIA and MICE) in terms of bias and mean squared error for mean
  estimation. This is due to the nonparametric nature of tree
  models. AMELIA and MICE are based on normality and multivariate
  linear model assumptions that are seldom satisfied in real complex
  data.   
\item Tree-modeling methods can adapt to available sample sizes.
  Conversely, parametric methods with a prespecified set of parameters
  to be estimated are problematic unless the sample size is
  substantially larger than the number of parameters.
\item Tree methods are not hindered or crippled by multi-collinearity
  or quasi-complete separation. In fact, collinearity is often used to
  advantage in tree algorithms; e.g., CART surrogate splits
  \citep{cart} and GUIDE linear splits \citep{guidec}.  
\item For cases that involve large numbers of candidate predictors,
  tree methods can be orders of magnitude faster compared to
  traditional methods, which are impracticable on large data sets.
  The speed advantage increases exponentially with number of
  variables. Therefore it is invaluable to imputation of multiple data
  sets, bootstrapping, and other variance estimation techniques in
  large surveys.
\end{enumerate}

In addition, note that the nodes in the tree models as displayed in
Figures~\ref{fig:GCT:630}--\ref{fig:RRT:630} can be interpreted as
nonresponse adjustment cells.  For example, \citet{LV03} extended the
ideas of \citet{little86} to outline two strategies for reducing the
number of adjustment cells: (a)~choosing cells that are homogeneous
with respect to the probability of response, and (b)~ choosing cells
that are homogeneous with respect to outcome variable $Y$.  They noted
that weighting based on either of these methods of grouping removes
non-response bias in estimating population means: The nodes produced
by the classification tree method constitute cells that are
homogeneous with respect to the estimated probability of response as
shown in Figures~\ref{fig:GCT:630} and \ref{fig:RCT:630}, while the
regression tree method produces cells that are homogeneous with
respect to the predicted outcome variable, INTRDVX, as shown in
Figures~\ref{fig:GRT:630} and \ref{fig:RRT:630}
 
This work here can be extended in several directions.  First, it would
be useful to consider design-adjusted versions of the procedures
proposed here.  This potentially would include the use of weights in
the growth of individual trees or forests; in estimation of tree
parameters; and in the corresponding estimation of (sub)population
means. Of special interest would be evaluation of the extent to which
a given procedure may be sensitive to specified patterns of
heterogeneity in the weights.  For example, as noted in
Section~\ref{sec:methods}, all tree and forest algorithms use
approximations, and the properties of the resulting procedures can be
sensitive to the extent to which a given dataset is consistent with
the approximations used for that procedure; and some
weight-heterogeneity patterns may exacerbate that sensitivity.  In
addition, several analyses here used some predictor variables that are
equal to membership indicators for certain large primary sample units.
It would be of interest to extend previous literature on the use of
stratum and PSU labels in regression to the current case.

Second, development of appropriate variance estimators would provide
important tools for use in pruning of trees, and for inference related
to tree- or forest-based mean estimators.  This would require tree-
and forest-related extensions of standard theorems on the properties
of replication-based variance estimators under complex sample designs.
Of special importance would be conditioning arguments arising from the
fact that the structure of a given tree is data-driven and not
determined a priori.

Third, the current paper focused on estimation of the means of the
``interest and dividend'' variable, and a related missing-data flag,
for the U.S. Consumer Expenditure Survey. Some users of CE data,
however, are interested in carrying out econometric analyses based on,
e.g., regression, related generalized linear models and more complex
hierarchical models.  For those situations, one may need to impute
simultaneously a substantial number of missing income and expenditure
variables for a given consumer unit, and evaluation criteria for the
properties of the resulting imputation procedure may be more complex,
as discussed in, e.g., \citet{Rubin96}.

\appendix

\section{Details of variables}
\label{sec:appendix}

Table~\ref{tab:19var} lists the 19 $X$ variables in the simulation
experiment where MICE did not fail.  Tables~\ref{tab:top:class} and
\ref{tab:top:reg} list the top predictors of INTRDVX\_ and INTRDVX
according to GUIDE.  Table~\ref{tab:others} gives definitions of five
additional $X$ variables that appear in the tree diagrams.
Table~\ref{tab:missinfo} gives the names of the 124 $X$ variables with
missing values (the number missing is beside name).
Tables~\ref{tab:psu}--\ref{tab:earncomp} give the value codes of the
categorical variables appearing in the trees in
Figures~\ref{fig:GCT:630}--\ref{fig:RRT:630}.

\begin{table}
  \centering
  \caption{Nineteen $X$ variables for which MICE did not fail repeatedly; 
    last column gives number of missing values}
  \label{tab:19var} \vspace{0.5em}
  \begin{tabular}{rlp{3.9in}r} \hline
    Rank & Name & Definition & \# \\ \hline
    1& AGE\_REF & Age of reference person & 0 \\
    2& BATHRMQ & Number complete bathrooms in unit & 21 \\
    3& BEDROOMQ & Number of bedrooms in unit & 25 \\
    4& BLS\_URBN & Urban or rural & 0 \\
    5& BUILDING & Which best describes this building? & 0 \\
    6& CUTENURE & Housing tenure (owned with or without mortgage,
                  rented, etc.; 6 unordered values & 0 \\                  
    7& EARNCOMP & Composition of earners (8 unordered values) & 0 \\
    8& EDUC\_REF& Education of reference person (9 ordered values) & 0 \\
    9 & FAM\_TYPE & CU type based on relationship of members to reference person;
                      ``own'' children include blood-related sons and
                      daughters, step children and adopted children & 0 \\
    10& MARITAL1 & Marital status of reference person (5 categories) & 0 \\
    11& NO\_EARNR& Number of earners & 0 \\
    12& NUM\_AUTO& Number of owned automobiles & 0 \\
    13& OCCUCOD1 & Occupation (18 categories) &1697\\
    14& REF\_RACE& Race of reference person (6 categories) & 0 \\
    15& REGION & Region of country (NE, MW, S, W) & 33 \\
    16& ROOMSQ & Number of rooms in unit & 30 \\
    17& SEX\_REF & Sex of reference person (2 categories) & 0 \\
    18& ST\_HOUS & Are these living quarters presently used as student housing by 
                   a college or university? (yes/no) & 0 \\
    19& ETOTA & Total outlays last and current quarters & 0 \\ \hline
  \end{tabular}
\end{table}

\begin{table}[htbp]
  \centering
  \caption{Top 20 predictors of INTRDVX\_ (flag variable for INTRDVX); 
    last column gives number of missing values out of 4609 records}
  \label{tab:top:class} \vspace{0.5em}
  \begin{tabular}{rlp{3.9in}r} \hline
    Rank & Name & Definition & \# \\ \hline
    1& RETS\_RVX & Flag variable for RETSURVX (amount received in retirement, survivor or disability pensions in past 12 months) & 0 \\
    2& FEDR\_NDX & Flag variable for Federal income tax refund & 0 \\
    3& STATE     & State (39 categorical values) & 486 \\
    4& LIQUIDX\_ & Flag variable for checking, savings, money market accounts, and CDs
                             & 0 \\
    5& SLRF\_NDX & Flag variable for refund state and local income tax refund & 0 \\
    6& PSU       & Primary sampling unit (21 categorical values) & 2579 \\
    7& IRAX\_    & Flag variable for value of all retirement accounts & 0 \\
    8& ERANKH    & Percent expenditure outlay rank & 367 \\
    9& INC\_RANK & Percent income rank & 367 \\
    10& SLOC\_AXX & Flag variable for state and local income taxes paid & 0 \\
    11& FEDRFNDX  & Amount of federal income tax refund & 2530 \\
    12& POV\_PY   & Is CU income below previous year's poverty
                    threshold? & 0 \\
    13& POV\_CY   & Is CU income below current year's poverty threshold? & 0 \\
    14& RETS\_RVB & Flag variable for range of retirement, survivor or 
                    disability pensions & 0 \\
    15& RETS\_VBX & Flag variable for median value of bracket range for RETSURVB
                    (retirement, survivor, or disability pensions) & 0 \\
    16& INC\_\_ANK& Flag variable for percent income rank & 0 \\
    17& ERANKH\_  & Flag variable for percent expenditure outlay rank & 0 \\
    18& RESPSTAT  & Completeness of income response (yes/no) & 0 \\
    19& ROYESTX\_ & Flag variable for income from royalty or estates and trusts &
                                                                                  0 \\
    20& FEDTAXX\_ & Flag variable for Federal income tax paid by all CU members &
    0 \\ \hline
  \end{tabular}
\end{table}

\begin{table}[htbp]
  \centering
  \caption{Top 20 predictors of INTRDVX;
    last column gives number of missing values out of 2838 records with non-missing 
    INTRDVX}
  \label{tab:top:reg} \vspace{0.5em} 
  \begin{tabular}{rlp{3.9in}r} \hline
    Rank & Name & Definition & \#NA \\ \hline
    1& AGE\_REF & Age of reference person & 0 \\
    2& CUTENURE & Housing tenure (owned with or without mortgage,
    rented, etc.; 6 unordered values & 0 \\
    3& STATE     & State (39 unordered values) & 316  \\
    4& RENTEQVX & Expected monthly rent of home, if rented out & 444 \\
    5& AGE2     & Age of spouse & 1201 \\
    6& INCNONW1 & Reason reference person did not work in past 12
    months (6 unordered values & 1869 \\
    7& STOCKX & Value of all directly-held stocks, bonds, and mutual funds 
                             & 2612 \\
    8& INCOMEY1 & Employer from which reference person received most
    earnings (6 unordered values) & 969 \\
    9& STOCKYRX & Median value of bracket range for STOCKX & 2630 \\
    10& FJSSDEDX & Estimated amount contributed to Social Security & 0 \\
    11& EARNCOMP & Composition of earners (8 unordered values) & 0 \\
    12& INC\_HRS1 & Number hours usually worked per week by reference person 
                             & 969 \\
    13& PERSOT64 & Number of persons over 64 in CU & 0 \\
    14& NO\_EARNR& Number of earners & 0 \\
    15& FRRETIRX & Social Security and Railroad Retirement income & 0 \\
    16& RENT\_QVX& Flag variable for RENTEQVX & 0 \\
    17& PROPTXPQ & Property taxes paid last quarter & 0 \\
    18& INC\_RANK& Percent income rank & 64 \\
    19& INCO\_EY1 & Flag variable for employer type & 0 \\
    20& INCN\_NW1 & Flag variable for reason did not work in past 12
    months & 0 \\ \hline
  \end{tabular}
\end{table}

\begin{table}[htbp]
  \centering
  \caption{Five variables appearing in
    Figures~\protect\ref{fig:GCT:630}--\ref{fig:RRT:630} but
    not listed in Tables~\protect\ref{tab:19var}, 
    \ref{tab:top:class} and \ref{tab:top:reg}; none has missing values}
  \label{tab:others} \vspace{0.5em} 
  \begin{tabular}{lp{4.1in}} \hline
    Name & Definition \\ \hline
    SMSASTAT  & Does CU reside inside a Metropolitan Statistical Area? (yes/no) \\
    TOTXEST & Estimated total taxes paid \\
    FFTAXOWE & Weighted estimate for federal tax liabilities at the tax unit
    level \\
    HEALTHPQ & Amount health care last quarter \\
    VEHQ & Number of owned vehicles \\
    \hline
  \end{tabular}
\end{table}

\begin{table}
  \caption{Variables and their numbers of missing values}
  \label{tab:missinfo} \vspace{0.5em}
  \centering
  \begin{tabular}{llll} \hline
AGE2 1879      & INCWEEK2 1879   & OTHASTBX 4589  & ROOMSQ 30     \\
APTMENT 4535   & INC\_HRS1 1697  & OTHASTX 4564   & ROYESTB 4570  \\
BATHRMQ 21     & INC\_HRS2 2832  & OTHFINX 4571   & ROYESTBX 4570 \\
BEDROOMQ 25    & INC\_RANK 367   & OTHLNYRB 4605  & ROYESTX 4364  \\
BUILT 585      & INTRDVX 1771    & OTHLNYRX 4559  & SEX2 1879     \\
CNTRALAC 1459  & IRAB 4432       & OTHLOAN 3424   & SLOCTAXX 3990 \\
CREDFINX 4282  & IRABX 4432      & OTHLONB 4606   & SLRFUNDX 3167 \\
CREDITB 4584   & IRAX 3853       & OTHLONBX 4606  & STATE 486     \\
CREDITBX 4584  & IRAYRB 4407     & OTHLONX 4555   & STCKYRBX 4531 \\
CREDITX 4233   & IRAYRBX 4407    & OTHLYRBX 4605  & STDNTYRB 4591 \\
CREDTYRX 4248  & IRAYRX 3899     & OTHREGB 4594   & STDNTYRX 4483 \\
CREDYRB 4573   & LIQDYRBX 4448   & OTHREGBX 4594  & STDTYRBX 4591 \\
CREDYRBX 4573  & LIQUDYRB 4448   & OTHREGX 4338   & STOCKB 4550   \\
DEFBENRP 3490  & LIQUDYRX 3876   & OTHRINCB 4603  & STOCKBX 4550  \\
DIRACC 154     & LIQUIDB 4481    & OTHRINCX 4483  & STOCKX 4319   \\
EDUCA2 1879    & LIQUIDBX 4481   & OTHSTYRB 4585  & STOCKYRB 4531 \\
EITC 1032      & LIQUIDX 3827    & OTHSTYRX 4572  & STOCKYRX 4347 \\
ERANKH 367     & LMPSUMBX 4600	 & OTHSYRBX 4585  & STUDFINX 4511 \\
FEDRFNDX 2530  & LUMPSUMB 4600	 & OTRINCBX 4609  & STUDNTB 4598  \\
FEDTAXX 3752   & LUMPSUMX 4378   & POPSIZE 33     & STUDNTBX 4598 \\
FMLPYYRX 4514  & MEALSPAY 9      & PORCH 997      & STUDNTX 4473  \\
FS\_MTHI 4560  & MISCTAXX 4520   & POV\_CY 378    & SWIMPOOL 4045 \\
HHID 4531      & MLPAYWKX 4514   & POV\_PY 378    & WELFAREX 4596 \\
HISP2 1879     & MLPYQWKS 4508   & PSU 2579       & WELFREBX 4609 \\
HLFBATHQ 23    & NETRENTB 4582   & RACE2 1879     & WHLFYRB 4564  \\
HORREF1 4448   & NETRENTX 4258   & REGION 33      & WHLFYRBX 4564 \\
HORREF2 4495   & NETRNTBX 4582   & RENTEQVX 660   & WHLFYRX 4444  \\
INCNONW1 2912  & OCCUCOD1 1697   & RETSRVBX 4542  & WHOLIFB 4571  \\
INCNONW2 3656  & OCCUCOD2 2832   & RETSURVB 4542  & WHOLIFBX 4571 \\
INCOMEY1 1697  & OFSTPARK 1160	 & RETSURVM 3289  & WHOLIFX 4428  \\
INCOMEY2 2832  & OTHASTB 4589    & RETSURVX 3520  & WINDOWAC 3977 \\
\hline
  \end{tabular}
\end{table}

\begin{table}
  \caption{PSU codes in the CE data; only ``A'' size PSUs are identified,
    other PSUs are coded as missing} 
  \label{tab:psu} \vspace{0.5em}
  \centering
  \begin{tabular}{cp{5in}} \hline
    1109 & New York, NY \\
    1110 & New York, Connecticut suburbs \\
    1111 & New Jersey suburbs \\
    1102 & Philadelphia – Wilmington – Atlantic City, PA – NJ – DE - MD \\
    1103 & Boston – Brockton – Nashua, MA – NH – ME CT \\
    1207 & Chicago – Gary – Kenosha, IL – IN - WI \\
    1208 & Detroit – Ann Arbor – Flint, MI \\
    1210 & Cleveland – Akron, OH \\
    1211 & Minneapolis – St. Paul, MN – WI \\
    1312 & Washington, DC – MD – VA – WV \\
    1313 & Baltimore, MD \\
    1316 & Dallas – Ft. Worth, TX \\
    1318 & Houston – Galveston – Brazoria, TX \\
    1319 & Atlanta, GA \\
    1320 & Miami – Ft. Lauderdale, FL \\
    1419 & Los Angeles – Orange, CA \\
    1420 & Los Angeles suburbs, CA \\
    1422 & San Francisco – Oakland – San Jose, CA \\
    1423 & Seattle – Tacoma – Bremerton, WA \\
    1424 & San Diego, CA \\
    1429 & Phoenix – Mesa, AZ \\
    \hline
  \end{tabular}
\end{table}

\begin{table}
  \caption{Codes for variable FAM\_TYPE. CU type is based on relationship of members to reference person. ``Own'' children include blood-related sons and
    daughters, step children and adopted children.}
  \label{tab:famtype} \vspace{0.5em}
  \centering
  \begin{tabular}{cp{5in}} \hline
    1 & Husband and wife (H/W) only \\
    2 & H/W, own children only, oldest child under 6 years old \\
    3 & H/W, own children only, oldest child 6 to 17 years old \\
    4 & H/W, own children only, oldest child over 17 years old \\
    5 & All other H/W CUs \\
    6 & One parent, male, own children only, at least one child age under 18 years old \\
    7 & One parent, female, own children only, at least one child age under 18 years old \\
    8 & Single persons \\
    9 & Other CUs \\ \hline
  \end{tabular}
\end{table}

\begin{table}
  \caption{Codes for variable BUILDING}
  \label{tab:building} \vspace{0.5em}
  \centering
  \begin{tabular}{cp{5in}} \hline 
    1 & Single family detached (detached
        structure with only one primary residence; however, the structure
        could include a
        rental unit(s) in the basement, attic, etc.) \\
    2 & Row or townhouse inner unit (2, 3 or 4 story structure with 2
        walls in common with other units and a private ground
        level entrance; it may have a rental unit as part of structure) \\
    3 & End row or end townhouse (one common wall) \\
    4 & Duplex (detached two unit structure with one common wall between the units) \\
    5 & 3-plex or 4-plex (3 or 4 unit structure with all units occupying the same level or levels) \\
    6 & Garden (a multi-unit structure, usually wider than it is high,
        having 2, 3, or possibly 4 floors; characteristically the
        units
        not only have common walls but are also stacked on top of one another) \\
    7 & High-rise (a multi-unit structure which has 4 or more floors) \\
    8 & Apartment or flat (a unit not described above; could be located in the basement, attic, second floor or over the garage of one \\
    \hline
  \end{tabular}
\end{table}

\begin{table}
  \caption{Codes for EARNCOMP (composition of earners)}
  \label{tab:earncomp} \vspace{0.5em}
  \centering
  \begin{tabular}{cl} \hline
    1 & Reference person only \\
    2 & Reference person and spouse \\
    3 & Reference person, spouse and others\\ 
    4 & Reference person and others \\
    5 & Spouse only \\
    6 & Spouse and others \\
    7 & Others only \\
    8 & No earners \\
    \hline
  \end{tabular}
\end{table}

\section*{Acknowledgments}
The authors thank Steve Henderson and Geoff Paulin for very productive
discussions of the U.S. Consumer Expenditure Survey and to Matthew
Blackwell for help with the AMELIA software. The views expressed in
this paper are those of the authors and do not necessarily reflect the
policies of the U.S. Bureau of Labor Statistics.

\bibliography{paper}
\end{document}